\documentclass{article}
\usepackage{spconf,amsmath,graphicx}
\usepackage{color}
\usepackage{hyperref}
\usepackage{subcaption}
\usepackage{multirow}
\usepackage{balance}
\usepackage{verbatim}
\usepackage{hyphenat}
\usepackage{spconf,amsmath,graphicx,url}
\usepackage{enumitem,booktabs,multirow,hyperref,setspace}

\usepackage{xcolor,colortbl}

\definecolor{Gray}{gray}{0.85}
\definecolor{LightCyan}{rgb}{0.88,1,1}

\usepackage[
backend=biber,
style=ieee,
maxbibnames=3,
maxcitenames=3,
doi=false,isbn=false,url=false,eprint=false
]{biblatex} 

\defbibheading{bibliography}[\refname]{}

\title{Learnable Nonlinear Compression for Robust Speaker Verification}
\name{Xuechen Liu{$^1{}^,{}^2$}, Md Sahidullah{$^2$}, Tomi Kinnunen{$^1$}}
\address{
  {$^1$}School of Computing, University of Eastern Finland, Joensuu, Finland\\
  {$^2$}Universit\'{e} de Lorraine, CNRS, Inria, LORIA, F-54000, Nancy, France}
\begin{document}
\ninept
\maketitle
\begin{abstract}
In this study, we focus on nonlinear compression methods in spectral features for speaker verification based on deep neural network. We consider different kinds of channel-dependent (CD) nonlinear compression methods optimized in a data-driven manner. Our methods are based on power nonlinearities and dynamic range compression (DRC). We also propose multi-regime (MR) design on the nonlinearities, at improving robustness. Results on VoxCeleb1 and VoxMovies data demonstrate improvements brought by proposed compression methods over both the commonly-used logarithm and their static counterparts, especially for ones based on power function. While CD generalization improves performance on VoxCeleb1, MR provides more robustness on VoxMovies, with a maximum relative equal error rate reduction of 21.6\%.
\end{abstract}
\section{Introduction}
\label{sec:intro}

\emph{Automatic speaker verification} (ASV) \cite{asv2015, dnn_asv2021} is the task of verifying a person's identity using his or her voice. Modern ASV systems consists of three main components: acoustic feature extractor, speaker embedding extractor, and back-end classifier. In recent years, substantial improvement has been achieved by using \emph{deep neural networks} (DNNs) to implement, especially the last two components. Concerning speaker embedding extractor, statistical models such as \emph{i-vectors} \cite{ivector} have been replaced by deep models such as \emph{x-vector} with \emph{time-delayed neural network (TDNN)} \cite{xvector2018}. As for the back-end, recent studies have replaced \emph{probabilistic linear discriminant analysis} (PLDA) \cite{plda} with neural approaches \cite{nplda2020}.

Concerning features, however, many ASV systems still use \emph{mel-frequency cepstral coefficients} (MFCCs) \cite{mfcc1980}, which are not specialized for ASV and neglects information such as phase and temporal characteristics \cite{phase1, phase2, cosphase, pncc}. Meanwhile, spectrograms are also widely used \cite{voxceleb1, voiceid, enh_asv1, enh_asv2}. There are multiple types of spectrograms such as raw one where no filter is applied \cite{voxceleb1, voxceleb2} and more widely-used spectral energies output with mel filters \cite{magneto, ecapa}. Even if the spectral representations are usually higher-dimensional (hence, more expressive) than MFCCs, problems of lacking specialization and missing information remain. 

There are also attempts to replace hand-crafted features with neural networks \cite{sincnet, learnable_filterbanks2018, rawnet2019, wav2spk}. However, such design may be hard to interpret. Moreover, many state-of-the-art DNN extractors are based on convolutional kernels, whose modeling capability on variabilities across different frequency (\emph{channel} or \emph{subband}) components have been questioned \cite{sincnet, sincnet2_2019}. These potential shortcomings motivate the idea of optimizing signal processing modules of feature extractor, including spectrogram-based features. Such topic has been addressed recently for audio representation learning \cite{data_driven_harmonics} and ASV \cite{learnable_mfcc2021}, but expanding and optimizing nonlinear compression module has received less attention.

This study, motivated by the above, addresses \emph{channel-dependent} (CD) nonlinear compression of spectrogram energies. This is realized, as presented in Section \ref{sec:dynacompress}, by expanding the nonlinearity from a channel-independent to channel-dependent operation. Similar ideas on mel spectrogram have been effective in keyword spotting \cite{pcen_2017}, audio classification \cite{pcen_dcase_2021}, and far-field speaker verification \cite{pcen_pcmn_asv}. However, to the best of our knowledge, it has not been applied to various nonlinear compression methods in the task we consider. 

Our main contributions are summarized in two folds: 1) We leverage the power of such channel-dependent setting by revisiting two established nonlinear compression methods that have been efficient in previous works and generalized them to be channel-dependent, namely power function and dynamic range compression; 2) In order to capture different level of variabilities and compromise instabilities during the joint optimization, we propose a \emph{multi-regime} (MR) design based on CD.

\section{Nonlinear Compression in Acoustic Feature Extraction}
\label{sec:dynacompress}
When using spectrogram to extract features, as illustrated in Fig.~\ref{fig:methods}, we typically apply logarithmic compression to spectral energies. However, logarithm has a singularity at zero. This problem is often addressed by adding a small positive offset: $\log(x + \mathrm{offset})$. Even if it avoids the singularity, the ad-hoc design still lacks specificity to a given task and has unpredictable impacts for different kinds of input \cite{pcen_2017}. We consider two alternative parameterized methods and further make them to be channel-dependent, which are described below.

\subsection{Power Function}
\label{ssec:root}
The concept of applying power nonlinearity to compress the signal amplitude is inspired by human-auditory processing \cite{human_auditory, pncc_2012}. By using $X$ and $Y$ to denote the input and output magnitude spectra respectively, power nonlinearity is expressed as:
\begin{align}
    Y[t,f] = X[t,f]^{1/\alpha},
\label{eq:powerlaw}
\end{align}
where $\alpha$ is known as \emph{temperature coefficient} for the compression. $t$ and $f$ are the time and channel indices, respectively, for spectrogram energies. Experimentally, two particular values of $\alpha$ have been popular in speech front-ends. The first one is $\alpha=3$, known as \emph{cube-root} \cite{rastaplp1990, multitaper_mfcc_plp2013, mhec}. The other one is $\alpha=15$, known as \emph{power-law} \cite{kim_pncc_thesis2010}. Setting higher values of $\alpha$ can provide better recognition performance in the presence of white noise, while lower values may be required for maintaining accuracy for cleaner speech \cite{pncc}.

\subsection{Dynamic Range Compression}
\label{ssec:drc}
Power nonlinearity neither addresses foreground-background noise nor other variations. These problems can be addressed by applying dynamic range compression (DRC). It was proposed and applied originally to far-field keyword spotting as part of \emph{per-channel energy normalization} (PCEN) \cite{pcen_2017, pcen_2018}. 
PCEN has also been applied recently in audio event detection \cite{pcen_sed_2021}. Using the same notation as above, the DRC operation is defined by:
\begin{align}
    Y[t,f] = (X[t,f] + \delta)^{r} - \delta^{r},
\label{eq:drc}
\end{align}
where $\delta > 0$ is a positive bias and $r$ is the exponential offset. This method bears resemblance to \emph{spectral subtraction} \cite{speech_restoration} in speech denoising. In the context of PCEN, it is applied to the spectral energies processed with \emph{automatic gain control} (AGC) \cite{agc}.. In this study, under the framework of DNN-based ASV, we investigate the efficiency of DRC by directly applying it to spectral energies. 

\begin{figure}
    \centering
    \includegraphics[width=0.5\textwidth]{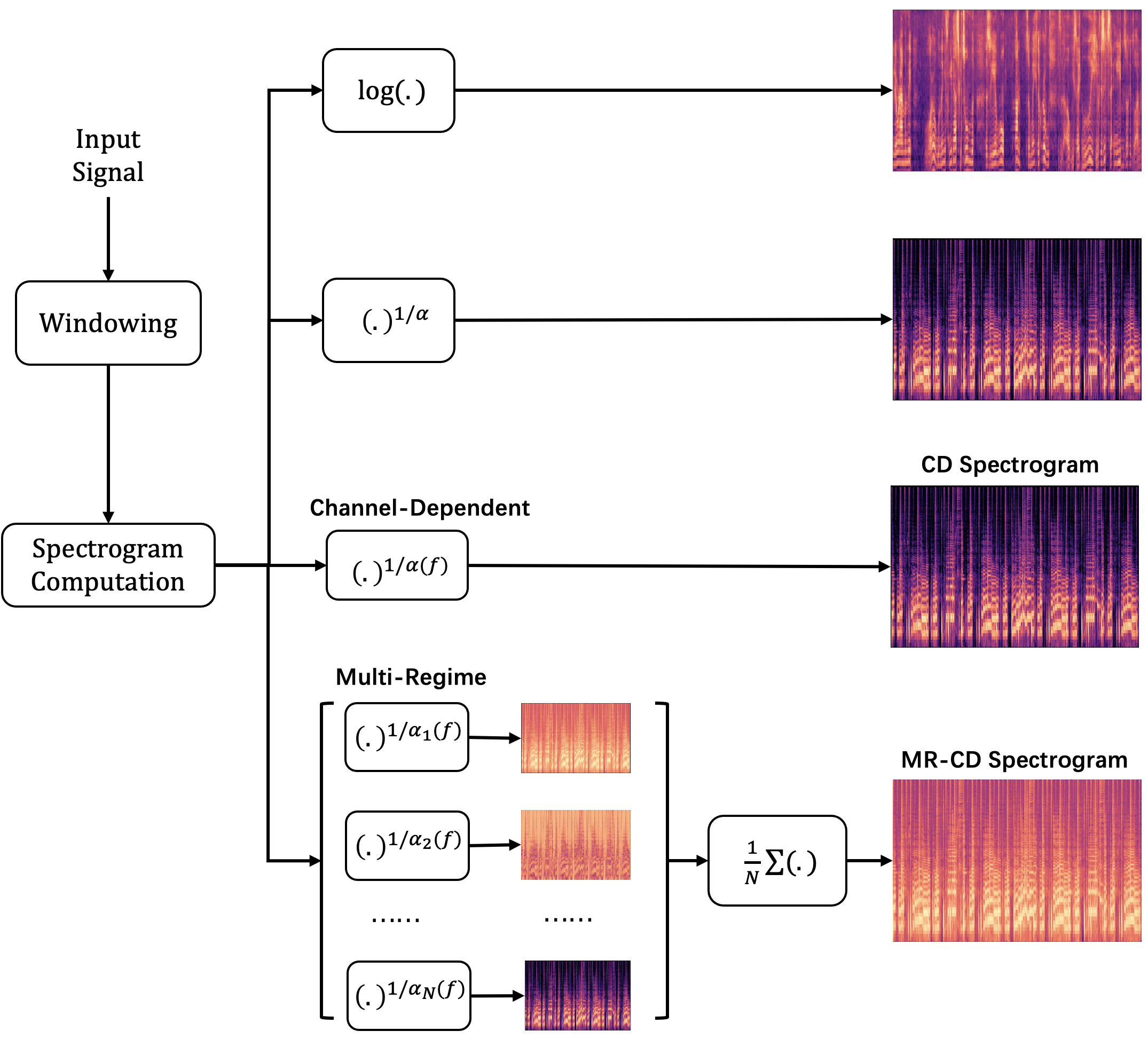}
    \caption{Feature extraction with log and proposed compression methods, using power function as example.}
\label{fig:methods}
\end{figure}

\section{Proposed Method}
\label{sec:methodology}

\subsection{Channel-Dependent Design}
\label{ssec:chandep}

While the control parameters in Eq. (\ref{eq:powerlaw}) and  (\ref{eq:drc}) can be set by hand, this may lead to suboptimal performance in a recognition task. Related prior studies on channel-dependent compression utilize information such as loudness and signal-to-noise ratio (SNR) \cite{snr_dependent_cd2}, which motivates data-driven settings. From the equations, we can see that the parameters $\alpha$, $\delta$, and $r$ are differentiable. Therefore, we propose to optimize them as part of the neural network by generalizing to their channel-dependent (CD) counterparts: $\boldsymbol{\alpha}=\alpha(f),\boldsymbol{\delta}=\delta(f), \boldsymbol{r}=r(f)$, where $f$ is the channel index. This design follows the proposal from \cite{pcen_2017}. The generalized parameters are then jointly optimized with the neural network. Furthermore, we employ \emph{kernelized initialization} where the parameters are initialized from their static counterparts \cite{learnable_mfcc2021}.

\subsection{Multi-Regime Design}
\label{ssec:multiregime}
Learnable parameters are tuned and selected by the training data during the learning process, thus might be suboptimal if domain mismatch between training and testing data is large, due to joint training which may let the parameters suffer from the overparameterized DNN models \cite{nips2020_ntk}. CD generalization of the parameters with kernel initialization may scrutinize such problem by larger search space and proper starting point, but it still may fail to have a wide-enough coverage of different level of speech variabilities. This is especially the case when the DNN has large number of layers, which may cause the problem of vanishing gradient when being back-propagated to first early layers, then nonlinearities \cite{training_dyna2020}.

Therefore, inspired by the design of \emph{multiple} feature maps in image processing \cite{image_processing} and audio event detection \cite{pcen_sed_2021}, we use a \emph{multi-regime} (MR) design by passing the spectrum to multiple submodules, with shared compression algorithm, but different initialized parameters, as shown in Fig.~\ref{fig:methods}. The output of each module is averaged to form the input for further operations. Using power function as an example:

\begin{align}
    Y[t,f] = \frac{1}{N}\sum_{i=1}^{N}(X[t,f])^{1/\alpha_{i}(f)}
\end{align}

We define the initial values by defining minimum and maximum to create $N$ evenly spaced values: $\alpha_{i}=\alpha_{\mathrm{min}} + (\alpha_{\mathrm{max}}-\alpha_\mathrm{min})*i/(N-1), i=1,...,N$, where $\alpha_{\mathrm{max}}$ and $\alpha_\mathrm{min}$ are maximum and minimum reference values and $N$ denotes number of intermediate spectrograms generated. The setup for this work is shown in Table \ref{tab:params}. For all cases in this work, $N=3$. Further tuning of number of intermediates and parameter search is left as future work.

\begin{table}[htbp]
    \centering
    \begin{tabular}{|c|c|c|}
        \hline
        Method & CD & MR-CD  \\ \hline
        \emph{cube root} \cite{rastaplp1990} & $\alpha=3$ & $\alpha_{\mathrm{max}}=3, \alpha_\mathrm{min}=1$ \\ \hline
        \emph{power law} \cite{mhec} & $\alpha=15$ & $\alpha_{\mathrm{max}}=15, \alpha_\mathrm{min}=1$ \\ \hline
        \multirow{2}{2em}{\emph{DRC}} & $\delta=2.0$ & $\delta_{\mathrm{max}}=2.0, \delta_\mathrm{min}=1.0$  \\
         & $r=0.5$ & $r_{\mathrm{max}}=1.0, r_\mathrm{min}=0.0$ \\ \hline
    \end{tabular}
    \caption{Parameter settings for kernel initialization. DRC values in CD are from \cite{pcen_2017} while for MR-CD it was hand-crafted based on \cite{pcen_2018} and pilot experiments.}
\label{tab:params}
\end{table}

\newcolumntype{g}{>{\columncolor{white}}c}
\begin{table*}[htbp]
  \renewcommand{\arraystretch}{1.1}
  \centering
  \begin{tabular}{|c|g|gg|gg|gggggg|}
  \hline
  & & \multicolumn{2}{|c|}{VoxCeleb1-E} & \multicolumn{2}{|c|}{VoxCeleb1-H} & \multicolumn{6}{|c|}{VoxMovies} \\ \hline
  Method & Design & EER(\%) & minDCF & EER(\%) & minDCF & E-1 & E-2 & E-3 & E-4 & E-5 & \textit{pooled} \\ \hline
  \rowcolor{Gray}
  \cellcolor{white}$\log$ & \cellcolor{white}- & 2.23 & 0.2676 & 4.43 & 0.5371 & 10.55 & 16.82 & 15.17 & 24.73 & 19.55 & 20.64 \\ \hline
  \rowcolor{Gray}
  \cellcolor{white}$\log(x + \mathrm{offset})$ & \cellcolor{white}- & 2.41 & 0.2920 & 4.93 & 0.6219 & 13.3 & 16.49 & 14.41 & 25.44 & 18.38 & 19.00 \\ \hline
  \rowcolor{Gray}
  \cellcolor{white}\multirow{3}{2.75em}{\emph{cube-root}} & \cellcolor{white}- & 1.93 & 0.2377 & 4.02 & 0.5090 & 10.68 & 14.74 & 13.00 & 25.21 & 16.61 & 18.21 \\ \cline{2-12}
  & CD & \textbf{1.91} & \textbf{0.2329} & \textbf{3.84} & \textbf{0.5071} & 10.84 & 14.79 & \textbf{12.93} & 25.71 & 16.67 & 18.33 \\ \cline{2-12}
  & MR-CD & 2.24 & 0.2876 & 4.52 & 0.5421 & 10.92 & \textbf{12.99} & 13.77 & \textbf{15.86} & 15.79 & \textbf{14.27} \\ \hline
  \rowcolor{Gray}
  \cellcolor{white}\multirow{3}{2.75em}{\emph{power-law}} & \cellcolor{white}- & 2.14 & 0.2542 & 4.38 & 0.5239 & 11.09 & 15.15 & 13.58 & 25.96 & 17.13 & 18.70 \\ \cline{2-12}
  & CD & 2.14 & 0.2505 & 4.29 & 0.5186 & \textbf{10.25} & 13.44 & 13.23 & 21.76 & 15.82 & 16.39 \\ \cline{2-12}
  & MR-CD & 2.78 & 0.3141 & 5.31 & 0.5683 & 12.07 & 14.27 & 15.35 & 18.05 & 16.92 & 15.80  \\ \hline
  \rowcolor{Gray}
  \cellcolor{white}\multirow{3}{2.75em}{\emph{DRC}} & \cellcolor{white}- & 2.25 & 0.2598 & 4.60 & 0.5629 & 11.09 & 14.07 & 14.14 & 22.20 & \textbf{15.16} & 16.89 \\ \cline{2-12}
  & CD & 2.67 & 0.2993 & 5.20 & 0.6408 & 11.95 & 14.07 & 14.45 & 19.78 & 16.24 & 16.38 \\ \cline{2-12}
  & MR-CD & 2.90 & 0.3526 & 5.81 & 0.6442 & 12.96 & 15.48 & 16.70 & 19.55 & 16.94 & 17.03 \\ \hline
  \end{tabular}
\caption{Speaker verification results on VoxCeleb and VoxMovies. `CD' means channel-dependent and `MR' means multi-regime setups. Rows with slight grey shades are regarded as baselines. `-' indicates the system with static nonlinearity without learning involved. For VoxMovies all results are reported in EER(\%).}
\label{tab:results}
\end{table*}

\section{Experimental Protocol}
\label{sec:experiments}
\textbf{Data}. For all experiments, we train the DNN speaker embedding extractor using the \emph{dev} set of VoxCeleb2 \cite{voxceleb2}, which consists of 5994 speakers. We report the performance of different methods on two evaluation sets: 1) The two test sets from VoxCeleb1 \cite{voxceleb1} following \cite{voxsrc2019}, known as \emph{VoxCeleb1-E} and \emph{VoxCeleb1-H}. 2) The recent \emph{VoxMovies} \cite{voxmovies}, which overlaps with VoxCeleb1 in terms of speakers and contains various levels of mismatch between the enrollment and test utterances. It consists of five trial sets, denoted E-1 (easiest) through E-5 (hardest). Besides condition-specific results, we also report the \emph{pooled} performance over the all five sets. 

\textbf{Features}. We use raw magnitude spectrogram obtained using \emph{short-time Fourier transform} (STFT) as the time-frequency representation, to which different compression methods are applied, as illustrated in Fig.~\ref{fig:methods}. The number of frequency bins $N_{\mathrm{STFT}}=512$ for all systems. The sampling rate is 16~kHz, and the STFT is computed using a 25~ms Hamming window every 10~ms.
Additionally, we include a system where the logarithm is factorized by an offset as part of the baseline, as mentioned in Section \ref{sec:dynacompress}: $\log(x + \mathrm{offset})$. The offset is parameterized by an exponential function $\mathrm{offset} = exp(\boldsymbol{\beta})$, where $\boldsymbol{\beta} = \beta(f)$ is CD and initialized with normal distribution.

\textbf{Speaker embeddings}. We use x-vector with \emph{extended TDNN} to generate speaker embeddings, following the design choice from \cite{Snyder_etdnn_2019} with two main modifications: 1) We replace the statistics pooling layer with attentive statistics pooling \cite{astats_pooling}; 2) Instead of multi-class cross-entropy, we use \emph{additive angular softmax} \cite{aam_softmax} as the training objective. We set the scaling factor $s=30$ and the margin $m=0.2$. We extract the embedding vectors from the first fully-connected layer after the pooling layer. The extracted vectors are centered and projected via a 150-dimensional \emph{linear discriminant classifier} (LDA).

\textbf{Evaluation}. For both VoxCeleb1 and VoxMovies, we train \emph{probabilistic LDA} (PLDA) classifier using VoxCeleb1. We report ASV performance in terms of \emph{equal error rate} (EER) and \emph{minimum detection cost function} (minDCF). For minDCF, the target speaker prior is $p_{\mathrm{tar}} = 0.01$ and detection costs were $C_\text{fa} = C_\text{miss} = 1.0$.

\section{Results}
\label{sec:results}

\subsection{Speaker Verification}
The results are presented in Table \ref{tab:results}. Let us first focus on VoxCeleb1. The EERs for both of the two power functions are improved from both the logarithm baseline and their static counterparts (marked as '-' in the `Design' column of the table) by the CD design. 
As part of the baseline, applying CD exponential offset on the logarithm compression degrades the performance for both test sets. The lowest EER on both test sets is obtained using \emph{cube-root} with CD, outperforming the baseline logarithm by 14.3\% and 13.3\%, respectively. This indicates the usefulness of CD. Nevertheless, the same design does not work well with DRC, which contradicts the findings reported in \cite{pcen_2017} with mel spectrogram (in a different task, though). This indicates that in ASV with spectrogram input, DRC may not combine well with CD compression.
For all compression methods, the MR-CD design degrades performance and fails to show substantial improvement over the logarithm. One reason could be suboptimal parameter initialization as in \cite{pcen_2017}, where the DRC parameters are set for far-field keyword spotting.

On the other hand, the trend is different for VoxMovies with more severe mismatch. Both \emph{pooled} and condition-specific indicate that improvements from CD generalization are modest, as opposed with the observations from VoxCeleb1. For \emph{cube-root}, CD actually degrades the performance for pooled and individual trial sets apart from E-3, where 14.7\% relative EER reduction is obtained over the logarithm. However, for \emph{power-law} CD improves upon its static counterpart, with lowest EER on E-1 across all systems.

Generalization via both MR and CD brings substantial improvements on nonlinearities based on power function. Lowest EER of pooled, E-2 and E-4 is obtained using \emph{cube-root} with MR and CD. Its pooled performance outperforms its static counterpart by relatively 21.6\%. We notice the same for \emph{power-law}, whose pooled performance with MR and CD outperform its static version by relatively 15.5\%. This indicates usefulness of MR in enhancing the robustness. 

Nevertheless, the behavior of DRC is different from the power function. While its static setting reduced the EER from the logarithm by relative 18.2\%, applying CD results in only slight relative EER reduction (3.1\%) on pooled results and does not lead to better performance for the individual trial sets apart from E-4. Generalizing it with MR degrades the performance, which agrees with our observations for VoxCeleb. However, its static setting reaches the best performance across all methods. This indicates the parameters for DRC (bias and offset) being not suitable to cope with CD and MR, at least not within the DRC framework itself. Further investigation is needed on its parameterization.

\subsection{Representation Analysis}

\begin{figure}
    \centering
    \includegraphics[width=0.5\textwidth]{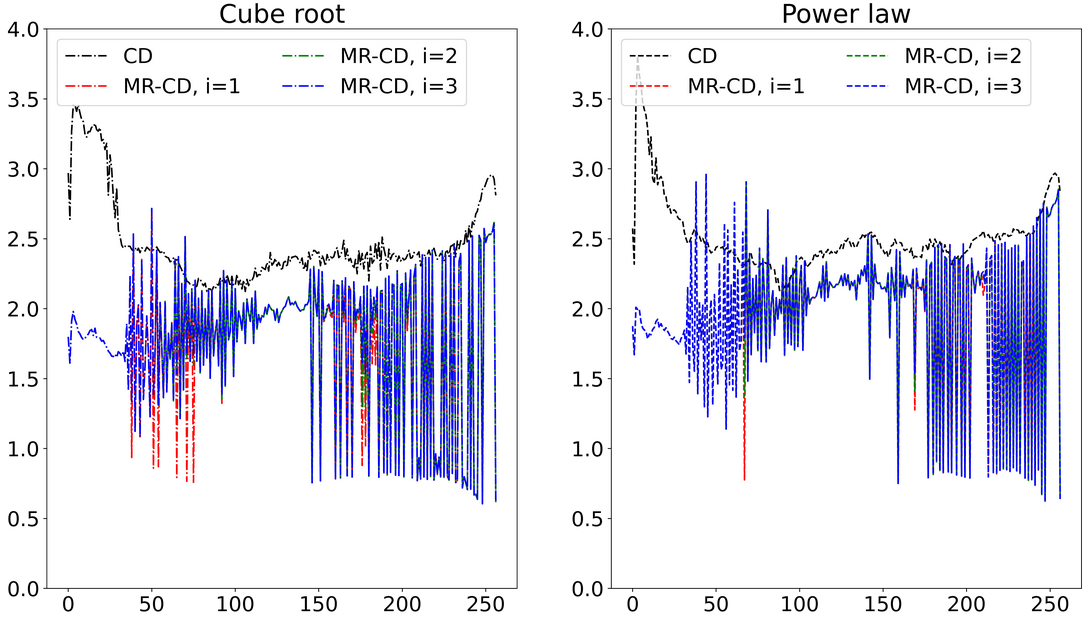}
    \caption{Learnt temperature values of nonlinear compression based on power function. The x-axis denotes frequency (channel) bin index and y-axis measures the values.}
\label{fig:representations}
\end{figure}

We illustrate the learnt temperature representation of the two power nonlinear functions from VoxCeleb2 \emph{dev} set (as described in Section \ref{sec:experiments}) in Fig.~\ref{fig:representations}. Note that for power function, larger magnitude of temperature parameters will result in higher compressing effect. 

As shown in the figure, applying only CD on power functions casts more compression on both low and high frequency regions (higher temperature values imply more aggressive compression, according to Eq.(\ref{eq:powerlaw})). Meanwhile, applying MR results in relatively less compression on some of the middle frequency components as well as low frequency regions, while relatively maintaining its pattern on high frequency components. Interestingly, we see that both power function methods result in similar numerical range, even if their initialized values are very different (Table~\ref{tab:params}). 

\section{Conclusion}
\label{sec:conclusion}
In this work, we have investigated the alternative nonlinearities for spectrogram compression and their dynamic, channel-dependent variants. We have extended their representation via channel-wise manner and utilized a multi-regime design based on it. Initialization on relevant parameters has been based on the corresponding static known values. We have evaluated the performance of proposed extended dynamic compression methods  for different degree of mismatch conditions. Results demonstrates the efficacy of the proposed methods on power nonlinearities, with a maximum of 21.6\% pooled EER reduction on VoxMovies. Future work may focus on: 1) extending the framework with other types of spectrogram; 2) exploring more advanced design and appropriate initialization and tuning methods, especially for DRC.



\section{Acknowledgements}
This work was partially supported by Inria Nancy Grand Est.


\section{References}
{
\printbibliography

@ARTICLE{asv2015,
  author={Hansen, J.H.L. and Hasan, T.},
  journal={IEEE Sig. Proc. Mag.}, 
  title={Speaker Recognition by Machines and Humans: A tutorial review}, 
  year={2015},
  volume={32},
  number={6},
  pages={74-99},
  doi={10.1109/MSP.2015.2462851}
}

@inproceedings{training_dyna2020,
  author={Erfan Loweimi and Peter Bell and Steve Renals},
  title={On the Robustness and Training Dynamics of Raw Waveform Models},
  year=2020,
  booktitle={Proc. Interspeech},
  pages={1001--1005},
  doi={10.21437/Interspeech.2020-17}
}

@INPROCEEDINGS{cosphase,
  author={Wu, Zhizheng and others},
  booktitle={Proc. ICASSP}, 
  title={Synthetic speech detection using temporal modulation feature}, 
  year={2013},
  volume={},
  number={},
  pages={7234-7238},
  doi={10.1109/ICASSP.2013.6639067}}

@article{rawnet2019,
  title={RawNet: Advanced end-to-end deep neural network using raw waveforms for text-independent speaker verification},
  author={Jung, Jee-weon and Heo, Hee-soo and Kim, ju-ho and Shim, Hye-jin and Yu, Ha-jin},
  journal={Proc. Interspeech},
  pages={1268--1272},
  year={2019}
}

@inproceedings{magneto,
  author={Daniel Garcia-Romero and Greg Sell and Alan Mccree},
  title={{MagNetO}: X-vector Magnitude Estimation Network plus Offset for Improved Speaker Recognition},
  year=2020,
  booktitle={Proc. Odyssey 2020 The Speaker and Language Recognition Workshop},
  pages={1--8},
  doi={10.21437/Odyssey.2020-1},
  url={http://dx.doi.org/10.21437/Odyssey.2020-1}
}

@inproceedings{ecapa,
  author={Brecht Desplanques and others},
  title={{ECAPA-TDNN}: Emphasized Channel Attention, Propagation and Aggregation in {TDNN} Based Speaker Verification},
  year=2020,
  booktitle={Proc. Interspeech},
  pages={3830--3834},
  doi={10.21437/Interspeech.2020-2650}
}

@inproceedings{sincnet2_2019,
  author={Erfan Loweimi and Peter Bell and Steve Renals},
  title={On Learning Interpretable {CNNs} with Parametric Modulated Kernel-Based Filters},
  year=2019,
  booktitle={Proc. Interspeech},
  pages={3480--3484},
  doi={10.21437/Interspeech.2019-1257}
}

@inproceedings{nips2020_ntk,
 author = {Liu, Chaoyue and Zhu, Libin and Belkin, Misha},
 booktitle = {Advances in Neural Information Processing Systems},
 editor = {H. Larochelle and M. Ranzato and R. Hadsell and M. F. Balcan and H. Lin},
 pages = {15954--15964},
 publisher = {Curran Associates, Inc.},
 title = {On the linearity of large non-linear models: when and why the tangent kernel is constant},
 url = {https://proceedings.neurips.cc/paper/2020/file/b7ae8fecf15b8b6c3c69eceae636d203-Paper.pdf},
 volume = {33},
 year = {2020}
}

@article{dnn_asv2021,
    title = {Speaker recognition based on deep learning: An overview},
    journal = {Neural Networks},
    volume = {140},
    pages = {65-99},
    year = {2021},
    issn = {0893-6080},
    doi = {https://doi.org/10.1016/j.neunet.2021.03.004},
    url = {https://www.sciencedirect.com/science/article/pii/S0893608021000848},
    author = {Z. Bai and X. Zhang}
}

@ARTICLE{rastaplp1990,
  author={Hermansky, H. and Morgan, N.},
  journal={IEEE Transactions on Speech and Audio Processing}, 
  title={{RASTA} processing of speech}, 
  year={1994},
  volume={2},
  number={4},
  pages={578-589},
  doi={10.1109/89.326616}
}

@ARTICLE{ivector,
  author={Dehak, N. and Kenny, P. J. and Dehak, R. and Dumouchel, P. and Ouellet, P.},
  journal={IEEE Transactions on Audio, Speech, and Language Processing}, 
  title={Front-End Factor Analysis for Speaker Verification}, 
  year={2011},
  volume={19},
  number={4},
  pages={788-798},
  doi={10.1109/TASL.2010.2064307}
}

@INPROCEEDINGS{learnable_filterbanks2018,
  author={Zeghidour, N. and Usunier, N. and Kokkinos, I. and Schaiz, T. and Synnaeve, G. and Dupoux, E.},
  booktitle={Proc. ICASSP}, 
  title={Learning Filterbanks from Raw Speech for Phone Recognition}, 
  year={2018},
  volume={},
  number={},
  pages={5509-5513},
  doi={10.1109/ICASSP.2018.8462015}
}

@INPROCEEDINGS{pcen_pcmn_asv,
  author={Liu, X and Sahidullah, M and Kinnunen, T},
  booktitle={2021 IEEE Automatic Speech Recognition and Understanding Workshop (ASRU) (to appear)}, 
  title={Parameterized Channel Normalization for Far-field Deep Speaker Verification}, 
  year={2021},
  volume={},
  number={},
}

@INPROCEEDINGS{sincnet,
  author={Ravanelli, M. and Bengio, Y.},
  booktitle={2018 IEEE Spoken Language Technology Workshop (SLT)}, 
  title={Speaker Recognition from Raw Waveform with SincNet}, 
  year={2018},
  volume={},
  number={},
  pages={1021-1028},
  doi={10.1109/SLT.2018.8639585}
}

@inproceedings{wav2spk,
  author={W. Lin and M.W. Mak},
  title={{Wav2Spk}: A Simple {DNN} Architecture for Learning Speaker Embeddings from Waveforms},
  year=2020,
  booktitle={Proc. Interspeech},
  pages={3211--3215},
  doi={10.21437/Interspeech.2020-1287}
}

@INPROCEEDINGS{data_driven_harmonics,
  author={Won, M. and Chun, S. and Nieto, O. and Serrc, X.},
  booktitle={Proc. ICASSP}, 
  title={Data-Driven Harmonic Filters for Audio Representation Learning}, 
  year={2020},
  volume={},
  number={},
  pages={536-540},
  doi={10.1109/ICASSP40776.2020.9053669}}

@inproceedings{nplda2020,
  author={S. Ramoji and others},
  title={{NPLDA}: A Deep Neural PLDA Model for Speaker Verification},
  year=2020,
  booktitle={Proc. Odyssey 2020},
  pages={202--209},
  doi={10.21437/Odyssey.2020-29},
  url={http://dx.doi.org/10.21437/Odyssey.2020-29}
}

@Inbook{human_auditory,
    author="Stern, Richard M.
    and Acero, Alejandro
    and Liu, Fu-Hua
    and Ohshima, Yoshiaki",
    editor="Lee, Chin-Hui
    and Soong, Frank K.
    and Paliwal, Kuldip K.",
    title="Signal Processing for Robust Speech Recognition",
    bookTitle="Automatic Speech and Speaker Recognition: Advanced Topics",
    year="1996",
    publisher="Springer US",
    address="Boston, MA",
    pages="357--384",
}

@phdthesis{kim_pncc_thesis2010,
  author       = {C. Kim}, 
  title        = {Signal processing for robust speech recognition motivated by auditory processing},
  school       = {Carnegie Mellon University, School of Computer Science},
  year         = 2010
}

@article{multitaper_mfcc_plp2013,
title = {Multitaper {MFCC} and {PLP} features for speaker verification using i-vectors},
journal = {Speech Communication},
volume = {55},
number = {2},
pages = {237-251},
year = {2013},
issn = {0167-6393},
doi = {https://doi.org/10.1016/j.specom.2012.08.007},
url = {https://www.sciencedirect.com/science/article/pii/S0167639312000994},
author = {M. J. Alam and others}
}

@article{mhec,
title = {Mean Hilbert envelope coefficients ({MHEC}) for robust speaker and language identification},
journal = {Speech Communication},
volume = {72},
pages = {138-148},
year = {2015},
issn = {0167-6393},
doi = {https://doi.org/10.1016/j.specom.2015.04.005},
url = {https://www.sciencedirect.com/science/article/pii/S0167639315000618},
author = {Seyed Omid Sadjadi and John H.L. Hansen},
}

@book{agc,
  title={Automatic Gain Control: Techniques and Architectures for RF Receivers},
  author={J. Prez and others},
  publisher={Springer-Verlag New York},
  year={2011}
}

@INPROCEEDINGS{pcen_sed_2021,
  author={Ick, C. and McFee, B.},
  booktitle={Proc. ICASSP}, 
  title={Sound Event Detection in Urban Audio with Single and Multi-Rate {PCEN}}, 
  year={2021},
  volume={},
  number={},
  pages={880-884},
  doi={10.1109/ICASSP39728.2021.9414697}}

@INPROCEEDINGS{xvector2018,
  author={Snyder, D. and others},
  booktitle={Proc. ICASSP}, 
  title={X-Vectors: Robust {DNN} Embeddings for Speaker Recognition}, 
  year={2018},
  volume={},
  number={},
  pages={5329-5333},
  doi={10.1109/ICASSP.2018.8461375}}

@ARTICLE{mfcc1980,
  author={Davis, S. and Mermelstein, P.},
  journal={IEEE Transactions on Acoustics, Speech, and Signal Processing}, 
  title={Comparison of parametric representations for monosyllabic word recognition in continuously spoken sentences}, 
  year={1980},
  volume={28},
  number={4},
  pages={357-366},
  doi={10.1109/TASSP.1980.1163420}
}

@INPROCEEDINGS{pcen_dcase_2021,
  author={Ick, C. and McFee, B.},
  booktitle={Proc. ICASSP}, 
  title={Sound Event Detection in Urban Audio with Single and Multi-Rate Pcen},
  year={2021},
  volume={},
  number={},
  pages={880-884},
  doi={10.1109/ICASSP39728.2021.9414697}
}

@INPROCEEDINGS{pcen_2017,
  author={Wang, Y. and others},
  booktitle={Proc. ICASSP}, 
  title={Trainable frontend for robust and far-field keyword spotting}, 
  year={2017},
  volume={},
  number={},
  pages={5670-5674},
  doi={10.1109/ICASSP.2017.7953242}}

@ARTICLE{pncc,
  author={Kim, C. and Stern, R. },
  journal={IEEE/ACM Trans. on Audio, Speech, and Language Processing}, 
  title={Power-Normalized Cepstral Coefficients ({PNCC}) for Robust Speech Recognition}, 
  year={2016},
  volume={24},
  number={7},
  pages={1315-1329},
  doi={10.1109/TASLP.2016.2545928}}

@INPROCEEDINGS{speech_restoration,
  author={Porter, J. and Boll, S.},
  booktitle={Proc. ICASSP}, 
  title={Optimal estimators for spectral restoration of noisy speech}, 
  year={1984},
  volume={9},
  number={},
  pages={53-56},
  doi={10.1109/ICASSP.1984.1172545}}

@ARTICLE{pcen_2018,
  author={Lostanlen, V. and others},
  journal={IEEE Sig.
Pro. Lett.}, 
  title={Per-Channel Energy Normalization: Why and How}, 
  year={2019},
  volume={26},
  number={1},
  pages={39-43},
  doi={10.1109/LSP.2018.2878620}}

@INPROCEEDINGS{Snyder_etdnn_2019,
  author={D. {Snyder} and others},
  booktitle={Proc. ICASSP}, 
  title={Speaker Recognition for Multi-speaker Conversations Using X-vectors}, 
  year={2019},
  volume={},
  number={},
  pages={5796-5800},
  doi={10.1109/ICASSP.2019.8683760}}

@inproceedings{astats_pooling,
  author={K. Okabe and others},
  title={Attentive Statistics Pooling for Deep Speaker Embedding},
  year=2018,
  booktitle={Proc. Interspeech},
  pages={2252--2256},
  doi={10.21437/Interspeech.2018-993},
  url={http://dx.doi.org/10.21437/Interspeech.2018-993}
}

@ARTICLE{aam_softmax,
  author={Wang, F. and Cheng, J. and Liu, W. and Liu, H.},
  journal={IEEE Signal Processing Letters}, 
  title={Additive Margin Softmax for Face Verification}, 
  year={2018},
  volume={25},
  number={7},
  pages={926-930},
  doi={10.1109/LSP.2018.2822810}}

@INPROCEEDINGS{learnable_mfcc2021,
  author={Liu, X. and Sahidullah, M. and Kinnunen, T.},
  booktitle={2021 IEEE International Symposium on Circuits and Systems (ISCAS)}, 
  title={Learnable {MFCC}s for Speaker Verification}, 
  year={2021},
  volume={},
  number={},
  pages={1-5},
  doi={10.1109/ISCAS51556.2021.9401593}}

@inproceedings{voxceleb2,
  author={J. Chung and others},
  title={VoxCeleb2: Deep Speaker Recognition},
  year=2018,
  booktitle={Proc. Interspeech},
  pages={1086--1090},
  doi={10.21437/Interspeech.2018-1929},
  url={http://dx.doi.org/10.21437/Interspeech.2018-1929}
}

@inproceedings{voxceleb1,
  author={A. Nagrani and J. Chung and A. Zisserman},
  title={VoxCeleb: A Large-Scale Speaker Identification Dataset},
  year=2017,
  booktitle={Proc. Interspeech},
  pages={2616--2620},
  doi={10.21437/Interspeech.2017-950},
  url={http://dx.doi.org/10.21437/Interspeech.2017-950}
}

@inproceedings{voxsrc2019,
  author={J. Chung and
               A. Nagrani and
               E. Coto and
               W. Xie and
               M. McLaren and
               D. A. Reynolds and
               A. Zisserman},
  title={Vox{SRC} 2019: The first VoxCeleb Speaker Recognition Challenge},
  year=2019,
  booktitle={ISCA archive},
}

@INPROCEEDINGS{voxmovies,
  author={Brown, A. and Huh, J. and Nagrani, A. and Chung, J. and Zisserman, A.},
  booktitle={Proc. ICASSP}, 
  title={Playing a Part: Speaker Verification at the movies}, 
  year={2021},
  volume={},
  number={},
  pages={6174-6178},
  doi={10.1109/ICASSP39728.2021.9413815}
}

@INPROCEEDINGS{pncc_2012,
  author={Kim, C. and Stern, R. M.},
  booktitle={Proc. ICASSP}, 
  title={Power-Normalized Cepstral Coefficients ({PNCC}) for robust speech recognition}, 
  year={2012},
  volume={},
  number={},
  pages={4101-4104},
  doi={10.1109/ICASSP.2012.6288820}}

@InProceedings{plda,
    author="Ioffe, S.",
    title="Probabilistic Linear Discriminant Analysis",
    booktitle="ECCV 2006",
    year="2006",
    pages="531--542"
}

@book{image_processing,
  address = {Upper Saddle River, N.J.},
  author = {Gonzalez, Rafael C. and Woods, Richard E.},
  description = {Digital Image Processing (3rd Edition)},
  publisher = {Prentice Hall},
  refid = {137312858},
  title = {Digital image processing},
  url = {http://www.amazon.com/Digital-Image-Processing-3rd-Edition/dp/013168728X},
  year = 2008
}

@inproceedings{phase1,
  author={Erfan Loweimi and Jon Barker and Oscar Saz Torralba and Thomas Hain},
  title={Robust Source-Filter Separation of Speech Signal in the Phase Domain},
  year=2017,
  booktitle={Proc. Interspeech},
  pages={414--418},
  doi={10.21437/Interspeech.2017-210}
}

@article{phase2,
title = {Short-time phase spectrum in speech processing: A review and some experimental results},
journal = {Digital Signal Processing},
volume = {17},
number = {3},
pages = {578-616},
year = {2007},
issn = {1051-2004},
doi = {https://doi.org/10.1016/j.dsp.2006.06.007},
url = {https://www.sciencedirect.com/science/article/pii/S105120040600087X},
author = {Leigh D. Alsteris and Kuldip K. Paliwal},
}

@INPROCEEDINGS{enh_asv1,
  author={Zhang, Chunlei and others},
  booktitle={Proc. ICASSP}, 
  title={Towards Robust Speaker Verification with Target Speaker Enhancement}, 
  year={2021},
  volume={},
  number={},
  pages={6693-6697},
  doi={10.1109/ICASSP39728.2021.9414017}}

@INPROCEEDINGS{enh_asv2,
  author={Xie, Weidi and others},
  booktitle={Proc. ICASSP}, 
  title={Utterance-level Aggregation for Speaker Recognition in the Wild}, 
  year={2019},
  volume={},
  number={},
  pages={5791-5795},
  doi={10.1109/ICASSP.2019.8683120}
}

@inproceedings{voiceid,
  author={Suwon Shon and others},
  title={{VoiceID} Loss: Speech Enhancement for Speaker Verification},
  year=2019,
  booktitle={Proc. Interspeech},
  pages={2888--2892},
  doi={10.21437/Interspeech.2019-1496}
}

@inproceedings{snr_dependent_cd2,
  author={Babak Nasersharif and Ahmad Akbari},
  title={A framework for robust {MFCC} feature extraction using {SNR-dependent} compression of enhanced mel filter bank energies},
  year=2006,
  booktitle={Proc. Interspeech},
}
}

\end{document}